\documentclass[preprint,showpacs,preprintnumbers,amsmath,amssymb]{revtex4}

\usepackage{graphicx}
\usepackage{dcolumn}
\usepackage{bm}

\begin{document}

\title{Spin Symmetry of the Bilayer Graphene Groundstate}

\author{Frank Freitag}
\author{Markus Weiss}
\email{Markus.Weiss@unibas.ch}
\author{Romain Maurand}
\author{Jelena Trbovic}
\author{Christian Sch\"onenberger}
\affiliation{
Department of Physics, University of Basel, Klingelbergstrasse 56, CH-4056 Basel, Switzerland}

\date{\today}

\begin{abstract}
We show nonlinear transport experiments on clean, suspended bilayer graphene that reveal a gap in the density of states.
Looking at the evolution of the gap in magnetic fields of different orientation,
we find that the groundstate is a spin-ordered phase. Of the three possible gapped groundstates
that are predicted by theory for equal charge distribution between the layers, we can therefore exclude the quantum anomalous
Hall phase, leaving the layer antiferromagnet and the quantum spin Hall phase as the only possible gapped groundstates
for bilayer graphene.
\end{abstract}


\maketitle

The isolation of monolayer graphene \cite{Novoselov2004} has given a new twist to the research on two-dimensional
electron systems, because graphene as a zero-gap semiconductor with a pseudo-relativistic dispersion relation
shows fundamentally new effects such as Klein-tunneling \cite{Katsnelson2006} that do not occur in conventional
two-dimensional electron gases (2DEGs) with a finite bandgap and a parabolic dispersion. The presence of two atoms
in the graphene unit cell is described by a sublattice-pseudospin, which acquires a Berry phase of $\pi$ on closed k-space
trajectories that include one of the K-points, which are the corner points of the first Brioullin zone and also constitute the Fermi
surface of charge neutral graphene. One consequence of this Berry phase is an unusual Landau level spectrum that leads to
the occurence of quantum Hall plateaus at conductances of half integer multiples of the Landau level degeneracy 
\cite{Novoselov2005,Zhang2005}. As the pseudospin degree of freedom is present also in Bernal stacked graphene multilayers,
they can be described by one low energy theory that explicitly includes the chiral nature of the charge carriers \cite{Min2008b}.  

Bilayer graphene in A-B stacking is a 2DEG with a parabolic dispersion for small energies $|E|\ll \gamma_1$, where
 $\gamma_1\approx$0.4 eV is the interlayer hopping parameter that links two atoms sitting on top of each other 
\cite{McCann2006}. As a consequence of the parabolic dispersion there is a finite density of states at the charge neutrality point
 (CNP) which, together with the weak dispersion leads to strong electron-electron interactions that make bilayer graphene at the 
CNP unstable towards interaction induced symmetry breaking
\cite{Min2008a,Barlas2009,Zhang2010,Nandkishore2010a,Nandkishore2010b,Vafek2010,Lemonik2010}. 

The exact nature of the electronic groundstate of bilayer graphene at charge neutrality is being discussed intensely
at the moment, with numerous theoretical investigations suggesting a large number of
phases. In each of these phases some of the three discrete degrees of freedom in bilayer graphene (spin, valley and layer)
undergo a transition to a lower symmetry state. While it is clear that for large magnetic fields quantum Hall ferromagnetism
will occur \cite{Barlas2008}, and that for strong perpendicular electric fields a layer polarized state will form 
\cite{Min2008a,Jung2011}, the nature of the groundstate at B=0 and E$_{\perp}$=0 is less obvious.
Whereas some experiments have found a conductive groundstate \cite{Mayorov2011,Bao2012}, the majority of experimental results 
\cite{Feldman2009,Weitz2010,Velasco2012,Bao2012,Freitag2012a,Freitag2012b,vanElferen2012,Veligura2012} point to an
insulating groundstate, an observation that limits the number of possible phases to the ones that are bulk gapped.

In this communication we show experimental results that further elucidate the nature of the bilayer graphene groundstate.
We have performed nonlinear conductance measurements at the CNP of clean, suspended bilayer graphene samples as a function
of magnetic fields oriented perpendicular and parallel to the graphene layer. Our measurements show a gap,
that increases strongly only in a perpendicular magnetic field, whereas it stays constant to a high precision
in a magnetic field that is exactly parallel to the bilayer plane. Our results allow us to make a statement about the spin order
of the bilayer graphene groundstate and therefore to narrow down further the number of possible candidates
for the electronic groundstate of bilayer graphene.

\begin{figure}[htb]
\centering
\includegraphics{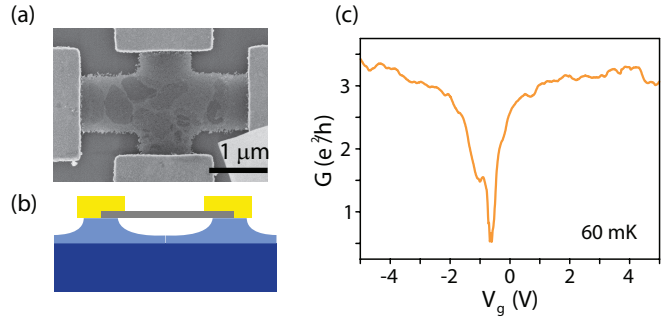}
\caption{a) SEM micrograph of the sample. b) Schematic of the sample structure.
c) Conductance G as a function of backgate voltage V$_{g}$.} \label{fig1}
\end{figure}

Bilayer graphene was exfoliated from natural graphite onto oxidized silicon wafers, and contacted with Cr/Au electrodes
using e-beam lithography. The bilayer was suspended by partly removing the oxide in a buffered hydrofluoric acid wet etch
and subsequent critical point drying, as described in \cite{Freitag2012b}. The data shown here were obtained on a
four-terminal sample in Hall-cross geometry that had been shaped by an Ar/O$_{2}$ plasma dry etching process (figure \ref{fig1}). 
We measured two-terminal conductance on contacts on opposite sides of the Hall-cross with the other two terminals floating
or connected to a high-impedance voltmeter for Hall measurements. All measurements were done in a
dip-stick dilution refrigerator at base temperature (T=60 mK).

Figure \ref{fig1} shows an SEM picture (a) and a schematic cross section (b) of the sample,
together with backgate characteristics of the device (c) measured at T=60 mK after current annealing \cite{Freitag2012b}.
The charge neutrality point (CNP) is visible as a deep minimum in the conductance G measured as a function of
gate voltage V$_{g}$ at -0.6 V, indicating the formation of an insulating state around charge neutrality.
The position of the CNP at a small negative gate voltage indicates a very small amount of residual dopants on the flake.

\begin{figure}[htb]
\centering
\includegraphics{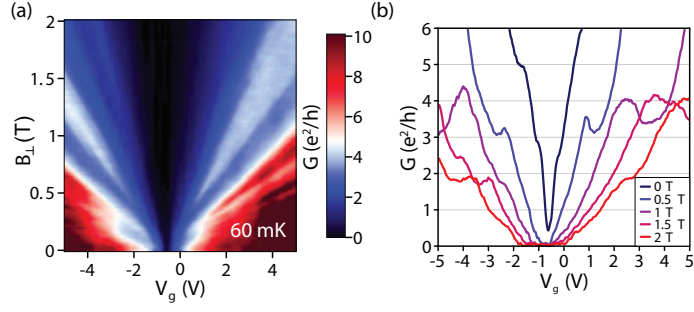}
\caption{a) Conductance G as a function of perpendicular magnetic field B$_{\perp}$ and gate voltage V$_{g}$.
b) Cuts through a) at several fixed magnetic fields.}\label{fig2}
\end{figure}

Figure \ref{fig2}a shows a measurement of the zero-bias conductance G as a function of gate voltage V$_g$ and
perpendicular magnetic field B$_{\perp}$. From the line cuts in figure \ref{fig2}b one can see that there is a
quantum Hall state at filling factor $\nu$=4 developed for B$_{\perp} \geq$ 1.0T, giving a lower limit of
10 000 cm$^2$/Vs for the charge carrier mobility. At higher fields additional intermediate plateaus start to occur,
indicating a lifting of the spin and valley degeneracies.

\begin{figure}[htb]
\centering
\includegraphics{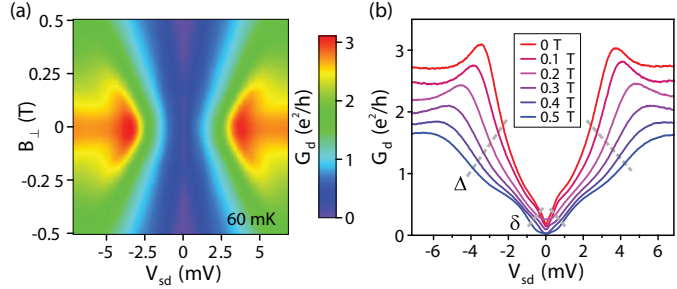}
\caption{a) Differential conductance G$_d$ at the CNP as a function of perpendicular magnetic field B$_{\perp}$
and bias voltage V$_{sd}$. b) Cuts through a) at several fixed magnetic fields.}\label{fig3}
\end{figure}

Differential conductance G$_d$=dI/dV at the CNP as a function of perpendicular magnetic field $B_{\perp}$ and
bias voltage V$_{sd}$ is shown in figure \ref{fig3}a. Several line-cuts at different magnetic fields are shown in
figure \ref{fig3}b. Similar to previous studies \cite{Velasco2012,Freitag2012a,Freitag2012b},
we find a strong suppression of conductance around zero bias, together with a BCS-like overshoot at a
finite voltage of V$_{sd}\approx$ 3.5 mV, indicating the formation of an interaction induced broken symmetry state
with a bulk gap.

The G$_d$(V$_{sd}$) curves in figure \ref{fig3}b show two inflection points that we associate with two different
gap energy scales $\delta$ and $\Delta$ \cite{Freitag2012a,Freitag2012b}.
Following the inflection points in finite perpendicular magnetic field, we find that the larger feature $\Delta$ increases
linearly with 3.1 meV/tesla, whereas the smaller feature $\delta$ shows a weaker response of 1.7 meV/tesla,
both of them however still much larger than the Zeeman energy for free electrons, which would amount to 0.116 meV/tesla
\footnote{Note that in our previous nomenclature\cite{Freitag2012a,Freitag2012b}, the present sample falls into category B2a.}.
As we have shown in previous studies \cite{Freitag2012a,Freitag2012b} the larger feature $\Delta$ only exists
in a narrow range of gate voltage around the CNP, whereas the small feature $\delta$ persists over the whole gate range that
is accessible with the present sample. We think that $\delta$ is due to localization in disordered areas on the sample perimeter,
and that $\Delta$ has to be associated with an interaction induced broken symmetry state located in the center region
of the sample. 

Interaction effects in bilayer graphene have been described with the so-called broken symmetry state quasiparticle Hamiltonian
\begin{equation}
\mathcal{H} = -\left(\frac{p^2}{2m^*}\right)\left[\cos\left(2 \phi_p\right)\sigma_x \pm \sin\left(2 \phi_p\right)\sigma_y\right] - 
\vec{\Delta}\cdot \vec{\sigma}
\label{eq1}
\end{equation}
with $\tan\left(\phi_p\right) = p_y/p_x$, $m^*$ the effective mass, $\vec{\sigma}$ the layer pseudospin vector,
+ and - chosen for valley K and K', respectively, and $\vec{\Delta}$ the order parameter of the
broken symmetry state \cite{Min2008a,Nandkishore2010a,Zhang2010,Zhang2011,Velasco2012}. 
For $\vec{\Delta} = (0,0,\Delta_z)$ the groundstate is gapped and equation (\ref{eq1}) predicts all electrons of the same spin and valley
(the same spin-valley flavour) to be located in the same layer, a state that has also been described as a layer
pseudospin magnet \cite{Min2008a}. Eight qualitatively different groundstates with distinct distributions of the four
spin-valley flavors across the two layers are possible. Three of them have no overall layer polarization,
and can be assumed to be energetically favorable in case of vanishing external electric fields. 
These three states are the quantum anomalous Hall state (QAH)\cite{Nandkishore2010b}, the layer antiferromagnetic state (LAF),
and the quantum spin Hall insulator (QSH). Their symmetries are determined by different order parameters in the quasiparticle
Hamiltionian (\ref{eq1}), namely $\Delta_z = \lambda \tau_z$ for the QAH, $\Delta_z = \lambda s_z$ for the LAF,
and $\Delta_z = \lambda \tau_z s_z$ for the QSH.
Here $\tau_z$ and $s_z$ represent the valley pseudospin and the electron spin, respectively. 
Due to a peculiar k-space topology \cite{Zhang2011}, each spin valley flavor has an intrinsic Hall conductivity,
the direction of which is given by $\tau_z \cdot$ sign($\Delta_z$): It depends on the valley and the sign of the order parameter
in equation (\ref{eq1}).
While for the LAF and QSH states the (charge) Hall conductivities of the four different spin-valleys cancel out,
in the QAH they add up, which means that for the QAH a nonzero quantized Hall conductance of 4e$^2$/h is expected
even in the absence of an external magnetic field. 
Although no net charge Hall conductivity is expected for the QSH phase, the resulting edge states will be helical,
leading to a quantized spin Hall effect.
More important, both QAH and QSH are expected to show a conductance of 4e$^2$/h in a two-terminal transport experiment,
in case the edge-states couple to the metallic contacts. However, as we have shown in a previous publication \cite{Freitag2012b},
this condition is not necessarily fulfilled in experiment, as the clean part of a sample that hosts the gapped phase can be separated
from the metallic contacts by a more disordered phase. This disordered phase will not let edge-states penetrate to the contacts,
but will confine them to isolated puddles in the sample center.
For the LAF charge, spin and valley Hall conductances cancel out, no edge-states can form, and consequently LAF is expected
to be fully insulating even in a two-terminal conductance measurement. Ignoring the effect of the magnetic field on the electron spins
the LAF state has also been predicted to show a gap whose size is independent of a perpendicular magnetic field,
whereas QAH and QSH should show a strong increase of the gap by 5.5 meV/tesla \cite{Velasco2012}. 
A theory including spin \cite{Kharitonov2012b} however predicts the formation of a canted antiferromagnet (CAF) that shows a
magnetic field dependence similar to that of QAH and QSH.

To identify the actual groundstate among the three candidates, it is useful to explore the three symmetries that might be preserved
in them, namely time reversal symmetry $\mathcal{T}$, spin rotation symmetry (SU$_2$), and valley Ising symmetry ($\mathcal{Z}_2$) 
\cite{Zhang2011,Velasco2012}. In each of the three phases, only one symmetry is preserved: time-reversal symmetry $\mathcal{T}$ in the QSH,
spin rotational symmetry SU$_2$ in the QAH, and valley Ising symmetry $\mathcal{Z}_2$ for the LAF.

The valley Ising symmetry $\mathcal{Z}_2$, a state being invariant under exchange of K and K', is broken in QAH and QSH,
but is difficult to assess in a transport experiment. Time reversal symmetry $\mathcal{T}$ is broken by the QAH and LAF,
and the case of QAH would manifest in a spontaneous Hall conductance at zero magnetic field,
and in general as an inequivalence of G(B) and G(-B). In a real world sample however,
puddles of different sign of the Hall conductance would exist next to each other,
so that a spontaneous Hall conductance would be unobservable. Spin rotation symmetry SU$_2$,
which is broken in the QSH and LAF,
with the spins in the bottom and top layer being locked into an antiferromagnetic arrangement,
can easily be assessed by looking at the response of the gap to a parallel magnetic field,
to exclude any orbital effects that are responsible for the big response to perpendicular field.
While LAF is a real antiferromagnet, with the spins in the top and bottom layer pointing into opposite directions,
the QSH also has to be considered a phase with antiferromagnetic spin order, however with the spin orientation
being opposite for different valleys. In both LAF and QSH, the electron spins are not free to move,
and a response to parallel magnetic field would be absent, as long as the Zeeman energy is smaller
than the exchange interaction responsible for the antiferromagnetic spin arrangement \cite{Zhang2012}.
In the QAH the electron spins are not ordered and would show a response to parallel field determined by the
Zeeman spitting of the spin up and spin down levels. This would lead to a decrease of the gap $\Delta$ by g$\mu_B$B.

\begin{figure}[htb]
\centering
\includegraphics{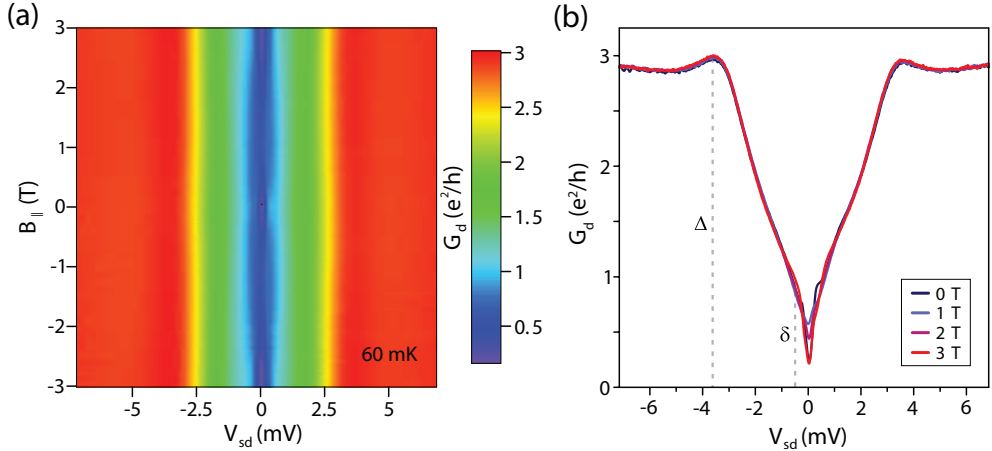}
\caption{a) G$_d$ at the CNP as a function of parallel magnetic field B$_{\parallel}$ and V$_{sd}$.
b) Cuts through a) at several fixed magnetic fields.}\label{fig4}
\end{figure}

In order to assess the response of $\Delta$  to a parallel magnetic field, we have performed nonlinear conductance measurements
in a vector magnet system that allowed adjusting the orientation of the magnetic field vector with high precision.
As the Zeeman splitting of the conduction electrons with g=2 is much smaller than the response of $\Delta$ to a perpendicular
magnetic field, the proper adjustment of the direction of B$_{\parallel}$ is crucial, and a small misalignment of a few degrees
will lead to a perpendicular field component that will change $\Delta$ much more than what is expected from Zeeman splitting alone.
Measurements were done with the sample mounted vertically, and B$_{\parallel}$ was applied in the direction of the bilayer plane.
As initial measurements showed an unreasonably large response of G$_d$(V$_{sd}$) to B$_{\parallel}$(see supplementary information),
we checked the correct alignment of the sample plane by applying an additional, small magnetic field in the horizontal direction, maximizing
the height of the BCS-like overshoot visible in G$_d$(V$_{sd}$) at V$_{sd}\approx$ 3.5 meV
(see figures \ref{fig3}b, \ref{fig4}b and supplementary information).
We found that the sample had a slight misalignment of $-$1.6$^o$ with respect to the vertical direction,
probably due to mechanical imperfections of the sample holder. Adding a small horizontal field component to the
vertical magnetic field, we were able to compensate for this misalignment and to adjust the magnetic field to the
sample plane with a precision of 0.1$^o$.
   
The results of a measurement of differential conductance at the CNP as a function of B$_{\parallel}$ and V$_{sd}$
are shown in figure \ref{fig4}.
Compared to the conductance as a function of perpendicular field (figure \ref{fig3}) it is clear from figures \ref{fig4}a and \ref{fig4}b
that the big gap $\Delta$ is not affected by a  parallel magnetic field. Following the outer inflection points
at V$_{sd}\approx$2.5 mV we find a magnetic field dependence of less than 60 $\mu$eV/tesla, which is significantly
smaller than the electron spin Zeeman splitting of 116 $\mu$eV/tesla (see supplementary information for details). 
Our experimental results therefore allow us to exclude a change of $\Delta$ due to the Zeeman splitting of the conduction
electrons. Note that a small dependence of G$_d$ on $B_{\parallel}$ occurs around zero bias V$_{sd}\approx$0. We think however
that this change in $G_d$ does not originate from the clean phase in the sample center and is not related to the gap $\Delta$.
According to the previous paragraph, this result shows that the electronic phase
that we have labeled $\Delta$ is a spin-ordered phase, where the electron spins are bound by some exchange interaction and
do not respond to an external magnetic field. Looking at the spin properties of the three possible candidates for $\Delta$,
we can conclude that $\Delta$ is not the quantum anomalous Hall (QAH) state, but that it has to be one of the remaining
two: the quantum spin Hall (QSH) or the layer antiferromagnet (LAF). To further distinguish QSH from LAF one could
make use of the spin Hall effect in the QSH, which should lead to a finite spin accumulation on the Hall terminals of a
Hall cross sample. Such a spin accumulation might be detectable via the inverse spin Hall effect, using a metal with strong
spin-orbit scattering, such as palladium, as contact material on the Hall terminals. Due to the non optimal homogeneity of
multi-terminal, current-annealed suspended graphene samples \cite{Freitag2012b}, the significance of such an experiment
would however have to be evaluated very carefully. 

In conclusion by performing nonlinear transport experiments on clean bilayer graphene for magnetic
fields of perpendicular and parallel orientation we have shown that the groundstate at charge neutrality is gapped,
and is a spin ordered phase. Of the three possible broken symmetry states that are suggested by theory, we can exclude that
a quantum anomalous Hall state is realized in bilayer graphene. The groundstate has to be either the quantum spin Hall state,
or the layer antiferromagnetic state. To distinguish the latter two from each other, further experimental work will be needed. 

\begin{acknowledgments}
We acknowledge financial support by the Swiss NCCR on Nanoscience and Nanotechnology, the NCCR on Quantum Science
and Technology, the ESF program Eurographene, the EU STREPS project SE2ND, and the Swiss National Science foundation.
\end{acknowledgments}

\end{document}


\title{Spin Symmetry of the Bilayer Graphene Groundstate\\
Supplementary Information}

\author{Frank Freitag}
\author{Markus Weiss}
\email{Markus.Weiss@unibas.ch}
\author{Romain Maurand}
\author{Jelena Trbovic}
\author{Christian Sch\"onenberger}
\affiliation{Department of Physics, University of Basel, Klingelbergstrasse 56, CH-4056 Basel, Switzerland}

\date{\today}
\maketitle

\section{Sample alignment procedure and data at finite angle}
\begin{figure}[htb]
\centering
\includegraphics[width=8cm]{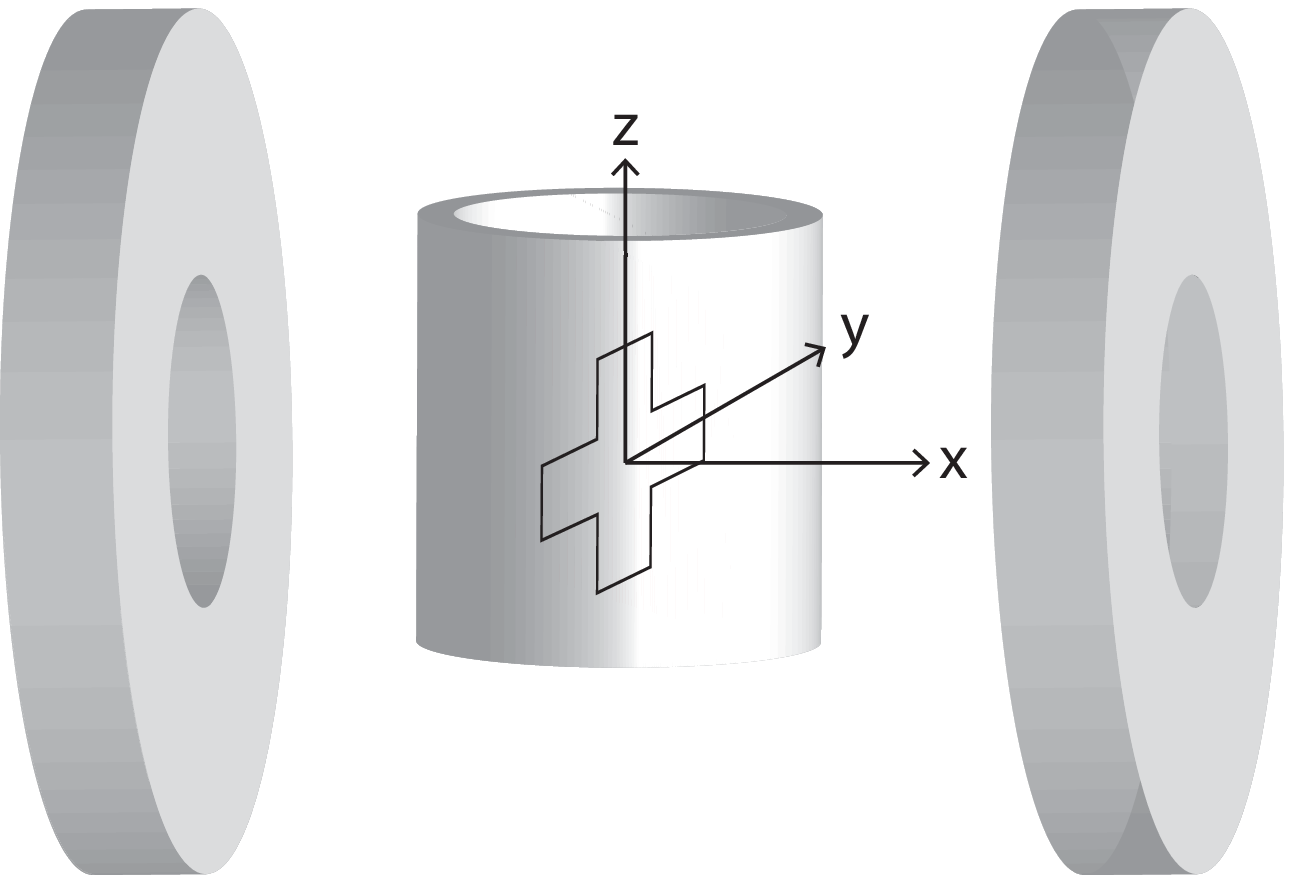}
\caption{Sketch of the vector magnet system and sample orientation.} \label{fig1}
\end{figure}

\begin{figure}[htb]
\centering
\includegraphics[width=18cm]{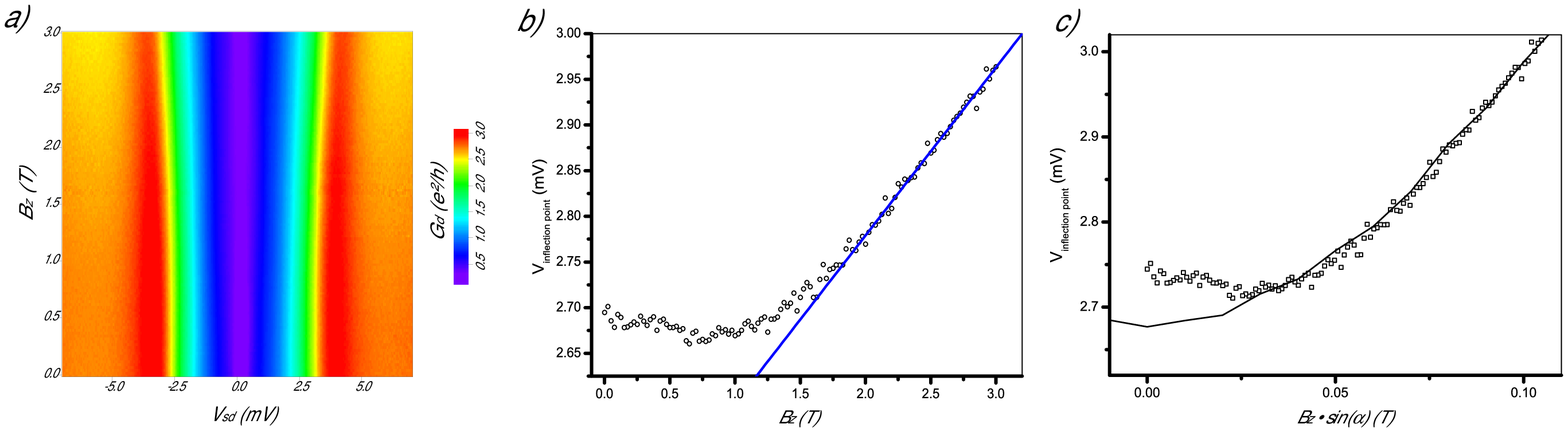}
\caption{a) Color scale plot of differential conductance as a function of source-drain voltage and B$_z$.
b) Position of the inflection point as a function of B$_z$. The solid line is a linear fit for 2T$\leq$B$_z$$\leq$3T.
c) The data from b) with the field axis rescaled by sin($\alpha$), with $\alpha$=2.0$^o$ (symbols).
The solid line corresponds to the position of the first inflection point for perpendicular field orientation
($\alpha$=90$^o$, see figure 3a in the main text).} \label{fig2}
\end{figure}

As stated in the main text, the precise parallel alignment of the sample plane to the magnetic field was crucial. Measurements were done
in a superconducting vector magnet system composed of a solenoid in z- and a split coil along the x-direction, as shown in figure \ref{fig1}.
The sample was cooled down in dip-stick dilution refrigerator, that could be rotated around the z-axis. It was mounted vertically
with the current contacts of the 4-terminal Hall cross roughly along the z-direction, and the Hall contacts
roughly along the y-direction. The sample plane was aligned parallel to y-direction very precisely by applying a field of 1 tesla along the
x-direction and rotating the dilution insert around the z-axis until the Hall signal was maximized. This was done at the charge neutrality point at elevated temperature (T$\approx$4.2K), a parameter regime where the Hall resistance was linear in B$_\perp$.
The result of initial measurements of G$_d$(V$_{sd}$,B$_z$) done after this first alignment procedure are shown in figure \ref{fig2}.\\
As stated in the main text, we take the outer inflection points at \textbar V$_{sd}$\textbar$\approx$2.5 mV as a measure for the gap $\Delta$.
Their position was determined by a numerical derivative of G$_d$(V$_{sd}$) and subsequent local fitting of the outer two extrema with a gaussian.
The average of the two inflection point positions at positive and negative bias voltage is shown in figure \ref{fig2}b.
A linear fit for large B$_z$ gives a magnetic field dependence of 183$\pm$4 $\mu$V/T, significantly more than what is expected
for Zeeman splitting (116$\mu$V/T). This magnetic field dependence can be explained by the presence of a finite perpendicular field component
that occurs due to a small tilt of the sample around the y-direction, probably due to mechanical imperfections of the sample holder. 
Rescaling the magnetic field axis by the sine of the effective angle between the sample plane and B$_z$, we can make the data
coincide with the corresponding data for perpendicular magnetic field orientation, assuming a finite angle of $\alpha$=2.0$^o$,
as shown in figure \ref{fig2}c. The deviations visible for B$<$30mT might be due to some small nonlinearity in the x-axis magnet,
that affected the measurement in perpendicular magnetic field.

To verify the misalignment angle, we applied a constant field of 1 tesla along the z-direction, and added a small field component along x, thereby slightly rotating the magnetic field vector around the y-axis. In order to find the perfectly parallel alignment of B we looked at
the BCS-like peak in G$_d$ at V$_{sd} \approx -$3.4 mV as a function of the small field component B$_x$ applied in addition to B$_z$=1T.
Given previous investigations of the magnetic field dependence of $\Delta$ \cite{Velasco2012}(see also figure 3a in the main text),
it is reasonable to assume that the peaks in G$_d$(V$_{sd}$) will be maximal for zero perpendicular field.

\begin{figure}[htb]
\centering
\includegraphics[width=18cm]{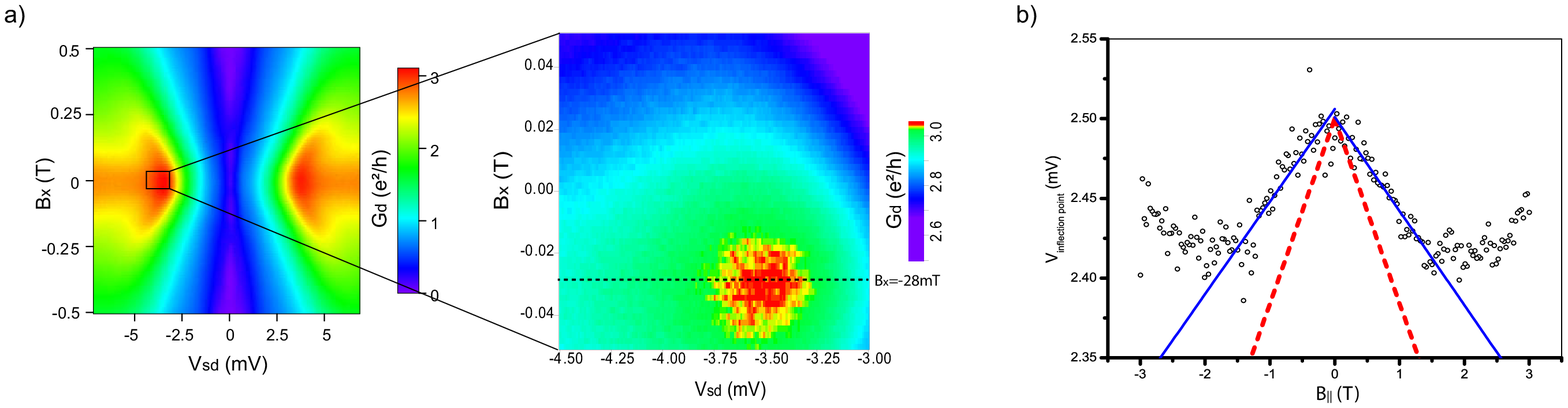}
\caption{a) G$_d$(V$_{sd}$,B$_x$) measured close to the "BCS"-like peak at V$_{sd}\approx -$3.5 mV with a constant
field B$_z$ applied. The maximum of G$_d$ occurs for B$_x$=$-$28 mT. The left colorscale plot is figure 3a from the main text.
b) Position of the inflection point extracted from the data shown in figure 4a of the main paper (symbols). 
The red dashed lines indicate the magnetic field dependence expected for Zeeman splitting of a spin with g=2 (116 $\mu$eV/T).
The blue solid lines are linear fits for 0$\leq$B$\leq$1.5T, giving a magnetic field dependence of 61$\pm$2 $\mu$eV/T.}\label{fig3}
\end{figure}

The results of such a measurement are shown in figure \ref{fig3}a.
We found that G$_d$ was maximized for B$_x$=$-$28mT applied in addition to B$_z$=1T, which corresponds to a rotation of the magnetic
field vector of -1.6$^o$ around the y-axis. The difference to the value of 2.0$^o$ deduced from the scaling analysis in figure \ref{fig2}c
could again be due to small nonlinearities in the x-axis magnet.
Note also that this procedure allowed us to find the exact parallel field alignment quickly while keeping the sample at T=60mK
and avoiding changes in G$_d$ due to temperature drift caused by eddy current heating,
that would be much more severe when sweeping the total magnetic field around zero.
For all further measurements in parallel magnetic field B$_z$ and B$_x$ were driven synchronously in order to maintain
the small tilt angle of -1.6$^o$ of the magnetic field vector. 

The data shown in figure 4a of the main text were measured this way, with no magnetic field dependence being visible to the naked eye.
To do a more thorough check, we determined the position of the inflection points by numerical derivation and local fitting,
as done for the data in figure \ref{fig2}. As can be seen in figure \ref{fig3}b, at small magnetic fields the inflection point
shifts to lower voltages linearly with about 60 $\mu$eV/T, a tendency that changes drastically at larger magnetic fields.
Although we currently do not know the reason for this non-monotonous behaviour, we can still exclude that this dispersion
is due to Zeeman splitting of the conduction electrons, because in this case a much stronger shift of 116 $\mu$V/T would be expected.

\section{Temperature dependence}

\begin{figure}[htb]
\centering
\includegraphics[width=10cm]{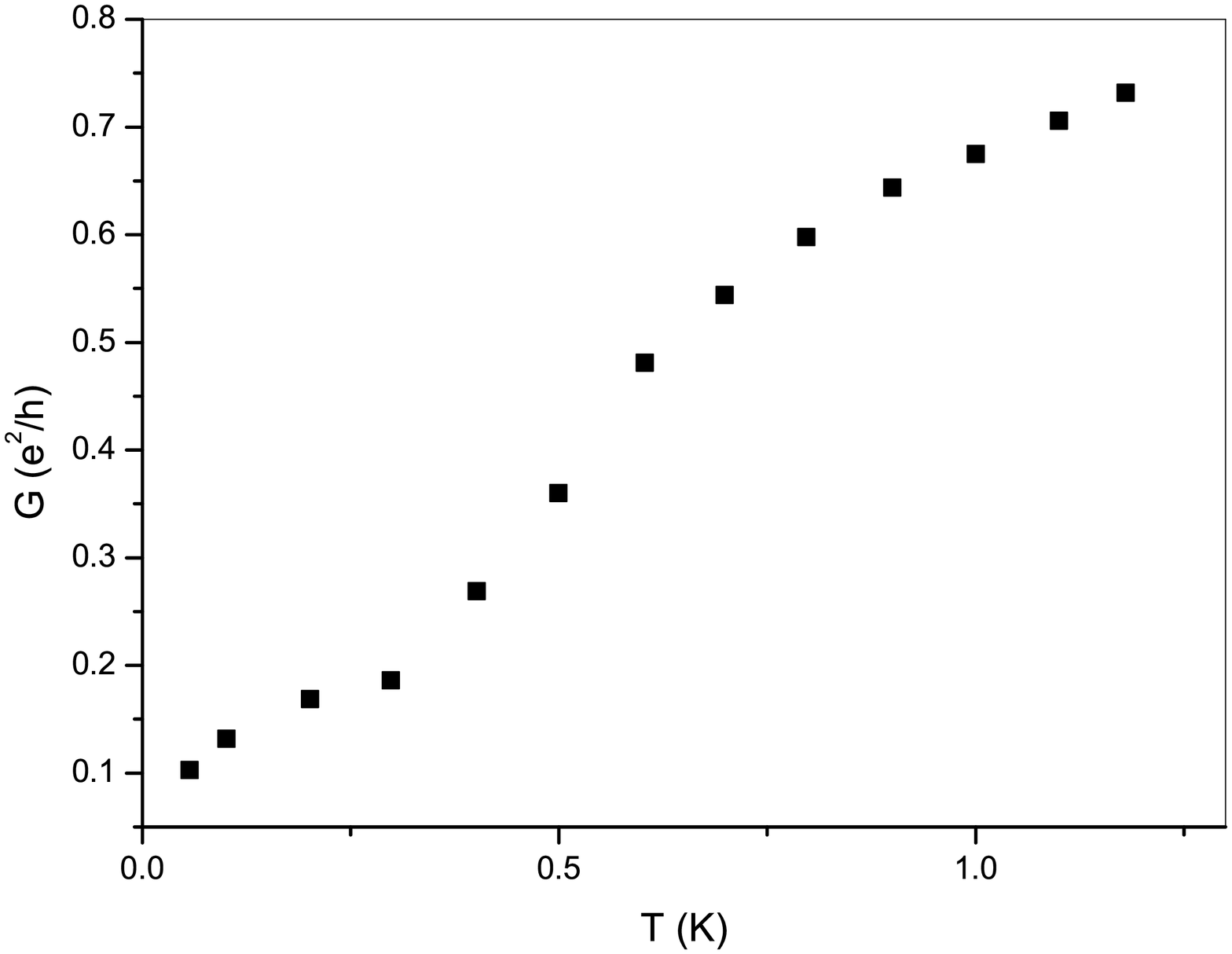}
\caption{Temperature dependence G(T) at B=0 at the CNP.}\label{fig4}
\end{figure}
The temperature dependence of the equilibrium conductance G(T) at the CNP is shown in figure \ref{fig4}. From 1.2K down to base temperature
the conductance decreases roughly by one order of magnitude, confirming the picture of a gapped groundstate in bilayer graphene.